# Oil Peak and the Decline of Net Energy: Policy Implications from an EROI–Entropy Perspective

Shunsuke Nakaya [1], Jun Matsushima [2]

[1] Systems Innovation, Faculty of Engineering, The University of Tokyo
Address: 7-3-1 Hongo, Bunkyo-ku, Tokyo, Japan

[2] Department of Environment Systems, Graduate School of Frontier Sciences, The University of Tokyo
Address: 5-1-5, Kashiwanoha, Kashiwa-shi, Chiba, 277-8563 Japan
ORCID: 0000-0003-1129-3964

Name and contact details of the corresponding author
Jun Matsushima (The University of Tokyo)
E-mail: jun-matsushima@edu.k.u-tokyo.ac.jp
Address: 5-1-5, Kashiwanoha, Kashiwa-shi, Chiba, 277-8563 Japan
TEL: +81-4-7136-4710

**Abstract:**

Net energy—the energy obtained from a resource after accounting for the energy expended in its acquisition—fundamentally determines the capacity of societies to sustain and expand. The extended Energy Return on Investment (EROIext) incorporates not only extraction and refining but also transport and end-use infrastructure, yet its long-term dynamics remain poorly understood. Here, we apply a Single-Cycle Lotka–Volterra (SCLV) model to the global petroleum system, calibrated with historical data from 1965–2012. The model projects trajectories of production, capital, and EROIext to 2100, and integrates an entropy-based indicator to evaluate the system's ability to maintain social order. Results show that oil production peaks around 2041, while EROIext declines continuously and falls below unity by 2081. This marks the point at which oil no longer delivers net energy, coinciding with the peak of capital stock, suggesting unsustainable investment in a diminishing resource. The rising entropy ratio signals declining systemic resilience. These findings underscore the importance of evaluating energy systems not only by quantity or cost but also by thermodynamic quality, with direct implications for balancing short-term energy security and long-term sustainability in policy design.

**Highlights:**

- System dynamics model linking extended EROI with entropy analysis
- Oil peak projected ~2041, capital peak lagging by four decades
- Declining net energy shows systemic energy–economy policy challenges

## 1. Introduction

Energy policy debates increasingly demand an understanding not only of the quantity and cost of energy resources but also of their thermodynamic quality and long-term viability. The ability of societies to sustain and expand their functions rests on net energy—the energy obtained from a resource after subtracting the energy required for its acquisition. When net energy is high, societies can allocate surplus energy to innovation, education, and governance; when it is low, resources must be reinvested into energy production itself, constraining economic and social progress (Odum, 1973; Tainter, 1988; Hall et al., 2009).

A widely used measure of net energy is the Energy Return on Investment (EROI), defined as the ratio of energy output to energy input (Cleveland et al., 1984; Hall et al., 1986). High-EROI fossil fuels such as coal and oil historically enabled industrialization and rising living standards (King, 2015). However, evidence shows that the EROI of oil has been declining over the long term, reflecting depletion of easily accessible reserves and the growing energy intensity of extraction (Gagnon, 2009; Hu, 2011; Brand-Correa et al., 2017). Among EROI variants, the extended form (EROIext) includes not only extraction and refining but also transport and end-use infrastructure, providing a more realistic boundary of the energy actually available to society (Hall et al., 1986).

Despite substantial progress in refining EROI metrics, two gaps remain. First, most studies focus on static or short-term accounting, offering limited insight into the long-term systemic dynamics of EROI when resource depletion, technological change, and capital investment interact. Second, the broader thermodynamic implications of declining EROI—particularly its relationship to entropy and societal stability—are rarely considered. While EROI measures energetic efficiency, it does not capture whether an energy system contributes to sustaining or undermining social order. Integrating entropy into the analysis provides such a perspective, as energy systems function by importing low-entropy resources and exporting high-entropy waste (Prigogine and Lefever, 1973; Hu et al., 2021).

Recent studies have begun to connect EROI with macro-level models of production and capital. Perissi et al. (2017, 2021) used Lotka–Volterra dynamics to describe overexploitation and peak behavior in resource systems, while Hu et al. (2021) interpreted EROI from an entropy perspective, framing energy systems as entropy-reducing structures. Yet these approaches remain only partially integrated, leaving the coupled long-term evolution of EROIext, production capacity, and entropy underexplored.

This study addresses that gap by applying a Single-Cycle Lotka–Volterra (SCLV) model to the global petroleum system. Petroleum resources and the capital required for extraction, refining, transport, and utilization are represented as coupled energy stocks. The model is calibrated with data from 1965 to 2012 and projected through 2100. By integrating entropy analysis, we assess not only the trajectory of EROIext but also the system's capacity to maintain social order by reducing societal entropy.

This integrated approach contributes in three ways. First, it treats oil production, capital investment, and EROIext as interdependent trajectories rather than isolated indicators. Second, it identifies critical thresholds, such as the year when EROIext falls below unity, after which oil ceases to be a net energy source. Third, it provides a thermodynamic interpretation of the societal value of petroleum by quantifying its entropy-reduction capability.

Ultimately, the findings have implications beyond theory. They underscore the importance of assessing energy systems not only by production volume and economic cost but also by their thermodynamic quality. In doing so, this study provides policy-relevant insights into how declining net energy and rising entropy may affect energy security, economic stability, and the design of sustainable energy transitions.

## 2. Methods

In this study, the petroleum energy production system is modeled using the Single-Cycle Lotka-Volterra (SCLV) model proposed by Perissi et al. (2021). Additionally, entropy is introduced into the SCLV framework based on the approach of Hu et al. (2021), and its dynamics are analyzed accordingly. This section first describes the basic structure of the SCLV model. It then explains how entropy is formulated using the parameters of the SCLV model. Finally, it details the data and fitting techniques used to model the petroleum energy production system.

### 2.1 SCLV model
#### 2.1.1 Lotka-Volterra (LV) model

The Lotka-Volterra model, originally developed by Lotka (1925) and Volterra (1927), describes the interactions between predators and prey using a system of coupled differential equations. Although the LV model is typically expressed in terms of population sizes, it can also be interpreted in terms of energy flows: energy is transferred from the prey ($L_1$) to the predator ($L_2$). Thus, in a general LV framework, $L_1$ and $L_2$ are treated as energy stocks, and the model describes the flow of energy between them (see Figure 1).

In the LV energy flow diagram, the prey ($L_1$ stock) receives energy from the external environment. This energy accumulates in $L_1$ and is gradually transferred to the predator ($L_2$ stock). During this transfer, a portion of energy is dissipated as waste heat to the environment. The energy that reaches $L_2$ also accumulates, but diminishes over time due to consumption and dissipation. For example, in a rabbit–fox ecosystem, rabbits gain energy from grazing (non-explicit environmental stock), and foxes gain energy by preying on rabbits. However, not all energy is transferred directly; some is lost due to hunting costs and metabolic processes (waste heat in Figure 1).

The energy exchange between the two stocks is represented by the following set of differential equations:

$$\frac{dL_1}{dt} = k_1 L_1 - k_2 L_1 L_2 \tag{1}$$

$$\frac{dL_2}{dt} = \eta k_2 L_1 L_2 - k_3 L_2 \tag{2}$$

The model incorporates four constants: $k_1$, representing the energy growth rate of the prey; $k_2$, the energy decay rate of the prey; $k_3$, the energy decay rate of the predator; and $\eta$, the energy transfer efficiency, which is the inverse of waste heat loss. While the Lotka–Volterra (LV) model is seldom employed in biological ecology due to its simplicity (Hall 1988), it has found application in economics, particularly in fisheries modeling (e.g., Smith, 1968; Perissi et al., 2021). In such contexts, the prey is conceptualized as a "resource," whereas the predator is regarded as the "capital" required to extract or utilize that resource—a perspective adopted in the present study.

### 2.1.2 Description of the SCLV model

In our model, the resource stock ($L_1$) corresponds to oil reserves, while the capital stock ($L_2$) represents the capital required to exploit these reserves, including the energy used for extraction, refining, transportation, and the infrastructure needed for petroleum utilization. Since fossil fuel resources such as oil regenerate at an extremely slow rate compared to the consumption and capital accumulation rates, the energy input rate from the environment ($k_1$) is assumed to be zero. The simplified Lotka-Volterra model under this condition is referred to as the Single-Cycle Lotka-Volterra (SCLV) model (Perissi et al 2021). The governing equations of the model are as follows:

$$\frac{dL_1}{dt} = - k_2 L_1 L_2 \tag{3}$$

$$\frac{dL_2}{dt} = \eta k_2 L_1 L_2 - k_3 L_2 \tag{4}$$

The model is characterized by three key parameters: $k_2$, representing the rate of conversion from resources to capital; $k_3$, denoting the degradation or depreciation rate of capital; and $\eta$, the efficiency of energy conversion. The term $dL_1/dt$ describes the rate of decline of the resource stock per unit time, which can be interpreted as the energy production rate.

The SCLV model has been used to describe production systems of slowly regenerating resources such as whales or gold (Perissi et al., 2021). Bardi and Lavacchi (2009) also applied it to petroleum systems by treating oil discoveries as resources and exploratory drilling as capital, demonstrating its relevance for oil production modeling. In this framework, EROI can be defined as the ratio of energy obtained by capital to the energy consumed by capital:

$$EROI = \frac{\eta k_2}{k_3} L_1 \qquad (5)$$

Eq. (5) expresses the energy return on investment (EROI) as a function of system efficiency ($\eta$), resource productivity ($k_2$), energy input requirements ($k_3$), and a scaling factor ($L_1$), thereby integrating both technical and resource-dependent parameters into a unified formulation. Because the capital stock in the model includes infrastructure required for petroleum utilization, the calculated EROI corresponds to EROIext.

## 2.2 Entropy and its integration with the model

Hu et al. (2021) approached the energy production system from a thermodynamic perspective using the theory of dissipative structures. This subsection explains the theoretical basis and formulation of entropy in our SCLV-based framework.

### 2.2.1 Energy production system and dissipative structure theory

Dissipative structure theory, established by Prigogine and Lefever (1973), incorporates open-system dynamics into the second law of thermodynamics. The law states that in isolated systems, entropy inevitably increases over time, leading to disorder and equilibrium. In contrast, open systems that exchange energy and matter with the environment may experience spontaneous emergence of ordered structures (dissipative structures) if the inflow of negative entropy surpasses a certain threshold. In open systems, energy dissipation does not simply lead to decay; rather, under continuous energy inflow, it can drive the emergence of new ordered structures that stabilize the system far from equilibrium. A system must meet several conditions to exhibit dissipative structures: it must be open, non-equilibrium, nonlinear, and highly sensitive to fluctuations (Xu et al., 2004). The entropy level must also be regulated, such that the total entropy change in the system is less than or equal to zero.

While originally a concept from physics, dissipative structures have been applied in ecology and social science. For example, Wang (2001) considered ecosystems—composed of biota and their environment—as dissipative structures that maintain order through exchanges of energy and matter. Likewise, socioeconomic systems, which are subsystems within ecosystems, consume low-entropy resources and expel high-entropy waste, thereby maintaining complex societal structures (Xu et al.., 2004). As an energy-intensive subsystem, the energy production system continuously exchanges energy and materials with both society and nature. It is inherently dynamic due to factors such as technological progress and new resource discoveries, and its behavior is shaped by nonlinear feedback mechanisms. Thus, it can be treated as a thermodynamic open system to which dissipative structure theory can be applied (Wang and Ouyang, 2012).

According to Hu et al. (2021), the energy production system increases internal entropy by consuming low-entropy inputs from the socioeconomic system. Meanwhile, it reduces external entropy by transforming natural resources into usable low-entropy energy, thereby providing a net reduction in societal entropy.

### 2.2.2 Linking EROI and Entropy

To discuss the relationship between EROI and entropy, Hu et al. (2021) developed an analytical diagram of an energy production system based on a Carnot heat engine (Figure 2). In Figure 2, the socio-economic system, the energy production system, and resources are regarded as a single system. Let $Q_1$ be the heat content of the materials and energy initially supplied to the energy resource development system, $S_1$ the corresponding entropy, $Q_2$ the heat contained in the waste heat and waste materials after the supplied materials and energy have been consumed, and $S_2$ the corresponding entropy. The useful work available for utilizing the energy resources is denoted as $W_1$. The heat content of the energy products is $Q_3$ and their entropy is $S_3$. When the energy products are transferred to and consumed in the socio-economic system, the useful work utilized for human purposes is $W_2$, while the waste heat and thermal energy contained in the waste are $Q_4$ and the corresponding entropy is $S_4$.

The total change in entropy of the system is given by

$$\Delta S = S_2 - S_1 + [-(S_4 - S_3)] \tag{6}$$

By multiplying this by the system temperature $T$, the entropy change is expressed in terms of heat. Then, applying the first law of thermodynamics yields

$$T\Delta S = W_1 - W_2 \tag{7}$$

If $W_1 = Q_1\alpha_1, W_2 = Q_3\alpha_2$ ($\alpha$ is the ratio of useful value to total energy), then

$$T\Delta S = Q_1\alpha_1 - Q_3\alpha_2 \tag{8}$$

If we simplify the EROI expression and assume that the energy quality coefficient is equal to the ratio of useful value to total energy, then

$$EROI = \frac{\alpha_2 Q_3}{\alpha_1 Q_1} \Leftrightarrow \alpha_2 Q_3 = EROI * \alpha_1 Q_1 \qquad (9)$$

Utilizing the relationship whereby the proportion of net energy equals the reciprocal of EROI, EROI is reformulated as:

$$EROI = \frac{-\Delta S + \Delta S_1}{\Delta S_1} \qquad (10)$$

where $\Delta S_1 = W_1/T$, and $W_1$ represents the net work invested $W_1$ for utilizing the energy resource, with $\Delta S_1$ denoting the change in environmental entropy when   is released as heat.

The greater the entropy reduction achieved by the energy production system, the larger the EROI becomes. Conversely, when the entropy change due to the energy production system is positive, the EROI falls below 1. Thus, within an energy production system considered as a thermodynamic open system, there exists a strong connection between EROI and entropy.

### 2.2.3 Integration with the SCLV model

By combining the above relationship between EROI and entropy with the relationship between EROI and other parameters in the SCLV model, entropy can be expressed in terms of the parameters of the SCLV model. As an assumption for this combination, it is presumed that the EROI derived from the entropy perspective, like the EROI derived from the SCLV model, is EROIext. The validity of this combination will be discussed in the Discussion section.

In the SCLV model, the EROI was given by

$$EROI = \frac{\eta k_2}{k_3} L_1 \qquad (11)$$

Combining this with the relationship between EROI and entropy yields:

$$\frac{-\Delta S + \Delta S_1}{\Delta S_1} = \frac{\eta k_2}{k_3} L_1 \qquad (12)$$

Rearranging gives:

$$\frac{\Delta S}{\Delta S_1} = 1 - \frac{\eta k_2}{k_3} L_1 \qquad (13)$$

Thus, the ratio of total entropy change to the change in environmental entropy when $W_1$ is released as heat ($\Delta S/\Delta S_1$) can be expressed in terms of the parameters of the SCLV model. The entropy ratio ($\Delta S/\Delta S_1$) can be interpreted as the expected entropy reduction effect when energy is supplied to a given system. For example, for a certain system, if $\Delta S/\Delta S_1 = 6.0$, and energy $x$ causes an entropy change of $s$ when released as heat, then if that energy is invested into the petroleum system, the total entropy decreases by $-6.0s$. The ratio $\Delta S/\Delta S_1$ can be regarded as an indicator of the efficiency of entropy reduction in a system and has characteristics similar to the EROI, which represents the efficiency of an energy production system. However, because entropy is more conceptual than EROI, the entropy ratio can be considered a value that accounts for a more holistic effect.

**2.3 Modeling the petroleum production system**

In this study, the SCLV model of the energy production system is applied to the case of petroleum production. The differential equations of the SCLV model are as follows:

$$\frac{dL_1}{dt} = -k_2 L_1 L_2 \tag{14}$$

$$\frac{dL_2}{dt} = \eta k_2 L_1 L_2 - k_3 L_2 \tag{15}$$

Here, $L_1$ represents resources, $L_2$ represents the capital required for the production, transportation, and utilization of energy, and $k_2$, $k_3$, and $\eta$ are constants. Modeling was conducted using data from 1965 to 2012. First, we describe how $L_1$ and $L_2$ were defined, and then explain how the constants were fitted using these values.

**2.3.1 Resource stock ($L_1$)**

Since the $L_1$ stock does not increase within the model, in this study $L_1$ was defined as the sum of the proved reserves, the probable reserves, and the expected additional reserves of crude oil in a given year (hereafter referred to as the "total recoverable reserves"). The proved reserves refer to the amount of crude oil confirmed to be recoverable with current technology; the probable reserves refer to the amount of crude oil expected to be discovered and recoverable with current technology; and the expected additional reserves refer to the increase in recoverable crude oil due to advances in recovery technology.

The annual total recoverable reserves from 1965 to 2012 were determined as follows. First, the total recoverable reserves in 1965 were calculated by subtracting the cumulative petroleum production up to 1965 from the ultimately recoverable reserves (cumulative production + proved reserves + probable reserves + expected additional reserves) published by the Japan Petroleum Development Association (2018). The cumulative production up to 1965 was obtained from the data presented in the paper by Court and Fizaine (2017). For the years after 1966, the total recoverable reserves for each year were defined as the total recoverable reserves in 1965 minus the cumulative production up to the previous year. Annual production data were taken from BP Statistical Review (BP 2002, 2010, 2021).

**2.3.2 Capital stock ($L_2$)**

The $L_2$ stock represents the capital required for producing, refining, transporting, and utilizing energy. The capital for production and refining was determined using the study by Court and Fizaine (2017), while the capital for transportation and utilization was estimated with reference to Hall et al. (2009). The methods for obtaining each are described below.

Court and Fizaine (2017) calculated the standard EROI (EROIst) for global oil production over the period 1860–2012 using macroeconomic data such as energy prices and GDP. In this study, we use the subset covering 1965–2012 (Figure 3). In line with conventional usage, EROIst is defined at the wellhead/field boundary, i.e., excluding refining, long-distance transportation, and end-use infrastructure. Accordingly, the input energy of EROIst covers exploration, drilling, lifting, and field operations, while refining and downstream processes are excluded. To estimate the energy requirements for production and refining, we divided EROIst values by annual petroleum production.

As with the $L_1$ stock, annual production values were taken from BP Statistical Review data (BP, 2021). (MMbbl = million barrels)

$$Input\ energy\ (MMbbl) = \frac{Production\ energy\ (MMbbl)}{EROIst} \quad (16)$$

Hall et al (2009), in considering the EROIext for the United States, referred to the energy required for transportation and utilization. According to Hall et al. (2009), when automobiles are assumed to be the primary means of energy transportation, the total of (1) the energy required for transportation and (2) the energy equivalent of the capital needed to use this transportation—such as vehicle maintenance costs and road maintenance costs—amounts to 64% of the energy produced domestically in the United States. In this study, we also assumed automobiles to be the primary means of energy transportation and further assumed that, in any given year, 64% of the energy produced was present as capital for transportation and utilization. This amount was included in the $L_2$ stock.

The energy flow diagram of the SCLV model developed in this study is shown in Figure 4. The process of energy transfer from the $L_1$ stock to the $L_2$ stock represents the movement of energy used for petroleum production as well as energy required to form the capital necessary for petroleum utilization. In the process of energy loss from the $L_2$ stock, the energy consumed in petroleum production and in maintaining the facilities required for petroleum production and utilization is released to the external system in forms that are unusable within the $L_2$ stock, such as heat. Energy that moves directly from the $L_1$ stock to the external system includes waste heat that could not be extracted as usable energy during production, as well as energy invested in sectors unrelated to the capital involved in petroleum production.

**2.3.3 Parameter fitting**

Modeling was performed using the above data. The methodology is described below. First, the SCLV model was discretized using a finite difference approximation. The discretized differential equations are as follows:

$$L_1(t+1) = (1 - \Delta t k_2 L_2(t))L_1(t) \tag{17}$$

$$L_2(t+1) = (1 + \Delta t \eta k_2 L_1(t) - \Delta t k_3)L_2(t) \tag{18}$$

Here, $L_1(t)$ and $L_2(t)$ represent the resource amount (MMbbl) and capital amount (MMbbl) in year $t$, respectively. The time step ($\Delta t$) was set to one year. Using this discretized SCLV model, parameter fitting was performed. The fitting was conducted using the least squares method. First, the squared errors between the resource and capital amounts calculated by the model and those obtained from the data were computed. Next, to match the order of magnitude between resource and capital values, the squared errors for resources and capital were each normalized by their respective maximum values.

Subsequently, the sum of the normalized squared errors was calculated:

$$\begin{aligned}Sum\ of\ squared\ errors \\ = \Sigma \frac{Squared\ error\ of\ resources}{Maximum\ squared\ error\ of\ resources} \\ + \Sigma \frac{Squared\ error\ of\ capital}{Maximum\ squared\ error\ of\ capit}\end{aligned} \tag{19}$$

Finally, the parameters ($k_2$, $k_3$, $\eta$) were fitted so as to minimize the sum of squared errors. The fitting was performed using the Generalized Reduced Gradient (GRG) method implemented in Excel Solver.

To assess the sensitivity of model outcomes to parameter variations, we vary each of $k_2$, $k_3$, and $\eta$ by ±10% and evaluate their effects on $L_1$, $L_2$, and production. Results are presented as confidence bands in the subsequent figures.

**3. Results**

The results are presented in three interrelated parts. First, the baseline data on resource and capital stocks are introduced to provide the empirical foundation of the analysis. Second, the outcomes of fitting the SCLV model to the observed data are reported, allowing for parameter estimation and model validation. Finally, the time-series evolution of EROIext and the entropy ratio, derived from the calibrated model, is analyzed to highlight their dynamic interplay over the study period.

### 3.1 Baseline data for resource and capital stocks

The "Original" blue solid line in Figure 5(a) shows the baseline data for resource stock ($L_1$), while the "Original" blue solid line in Figure 5(b) depicts the capital stock ($L_2$), and that in Figure 5(c) represents production ($dL_1/dt$). At the start of the observation period in 1965, the total recoverable oil reserves amounted to 4,586,591 million barrels. As capital for petroleum utilization increased, annual production grew, leading to a steady decline in reserves. By 2012, the total recoverable reserves had decreased to 3,469,680 million barrels.

### 3.2 Fitting Results of the SCLV Model

Using the baseline data, the parameters of the SCLV model were fitted, with the resulting values and total squared error summarized in Table 1. The modeled time evolution of the energy quantities for the resource stock, capital stock, and production is represented by the "Model" red solid lines in Figures 5(a), 5(b), and 5(c), respectively.

To assess the model's accuracy, the modeled values of $L_1$ (resource stock), $L_2$ (capital stock), and production ($dL_1/dt$) were compared with their corresponding baseline data. For resource stock and production, the discrepancy between the simulated and observed values tends to increase over time. In contrast, for capital stock, the difference is pronounced between 1965 and 1980, but the modeled values subsequently align closely with the baseline data. As shown in Figure 5(a), the model underestimates the depletion rate of $L_1$ relative to the actual data, resulting in a growing divergence over time. Figure 5(b) demonstrates that the difference in $L_2$ becomes negligible by 2012, while Figure 5(c) indicates that the deviation in production remains nearly constant after 1980.

The model was also employed to project the future dynamics of resource stock, capital stock, and production through 2100 (Figure 6). According to the projections, resource stock is expected to decline at an accelerating rate until 2043, when production reaches its peak; thereafter, as production decreases, the depletion rate of resource stock will slow. The capital stock capital stock is projected to grow rapidly from 1965 to 2022, reaching its maximum growth rate in 2022, after which growth will decelerate until the stock peaks in 2081, followed by a gradual decline. Production is anticipated to increase with accelerating growth from 1960 to 1996, after which growth slows, culminating in a peak in 2041 before entering a declining phase.

**3.3 Sensitivity Analysis**

A sensitivity analysis was conducted for the model developed in this study. The parameters $k_2$, $k_3$, and $\eta$ were individually varied by ±10%, and the resulting $L_1$ stock, $L_2$ stock, and production were calculated and plotted as error ranges (Figure 7).

According to Figures 7(a), 7(d), and 7(g), the $L_1$ stock shows low sensitivity to $\eta$, but its sensitivity to $k_2$ and $k_3$ increases over time. When the error rates in 2100 were calculated for variations in $k_2$, $k_3$, and $\eta$, the error rate for $\eta$ was 10%, whereas those for $k_2$ and $k_3$ were 47% and 43%, respectively. In terms of qualitative behavior, for variations in any of the coefficients, the rate of decline increased over time, but around 2040 the rate of decline began to decrease.

According to Figures 7(b), 7(e), and 7(h), the $L_2$ stock exhibits high sensitivity to $k_2$ from 2000 to 2080, after which the sensitivity decreases toward 2100. Sensitivity to $k_3$ and $\eta$, on the other hand, increases year by year. When the error rates in 2100 were calculated, variations in $k_2$, $k_3$, and $\eta$ yielded error rates of 7%, 27%, and 60%, respectively. Qualitatively, in all cases, the $L_2$ stock increases over time, reaches a peak, and then decreases. Changes in the timing of the peak position due to parameter variations are summarized in Table 2. A ±10% change in $k_2$, $k_3$, or $\eta$ resulted in a shift of approximately 10–20 years in the peak position.

According to Figures 7(c), 7(f), and 7(i), production shows gradually increasing sensitivity to $k_2$ from 1965 onward, but sensitivity decreases by 2050, after which it increases again. Sensitivity to $k_3$ increases annually, peaking around 2040 and then decreasing. Sensitivity to $\eta$ also increases annually but remains almost constant from around 2040 onward. When the error rates in 2100 were calculated, variations in $k_2$, $k_3$, and $\eta$ yielded error rates of 43%, 17%, and 57%, respectively. Qualitatively, similar to the capital stock, production increases over time, reaches a peak, and then decreases for variations in any of the parameters. Changes in the timing of the production peak due to parameter variations are summarized in Table 3. Compared with $\eta$ and $k_3$, $k_2$ exhibits higher sensitivity with respect to the peak position.

**3.4 Time evolution of EROIext and entropy ratio**

EROIext and the entropy ratio $\Delta S/\Delta S_1$ were computed from the fitted parameters using the following relationships:

$$EROIext = \frac{\eta k_2}{k_3} L_1 \tag{20}$$

$$\frac{\Delta S}{\Delta S_1} = 1 - \frac{\eta k_2}{k_3} L_1 \tag{21}$$

By substituting the fitted coefficients, the trajectories of EROIext (blue solid line in Figure 8) and the entropy ratio (ΔS/ΔS1) (red solid line in Figure 8) from 1965 to 2100 were calculated. In 1965, the EROIext was approximately 2.5, but it had decreased to about 2.0 by 2010. The decline continues thereafter, falling below 1.0 in 2081 and reaching around 0.79 by 2100. Regarding the entropy ratio, the results indicate a continuous increase from 1965 to 2100. The entropy ratio represents the expected entropy reduction effect when energy is introduced into a system, where a smaller value corresponds to a larger reduction effect. Therefore, it can be interpreted that the entropy reduction effect of the oil production system steadily diminishes over the period from 1965 to 2100. By 2081, the value becomes greater than zero, indicating that investing energy into the oil production system is expected to increase entropy rather than reduce it.

## 4. Discussion

In this section, we interpret the dynamics of the resource stock ($L_1$), capital stock ($L_2$), and production volume derived from the SCLV model. We also examine the relationships and interactions between these variables and EROIext. Additionally, we discuss the time lag between the peaks of capital and production, the implications of entropy ratio changes, and the broader societal and systemic implications.

### 4.1 Dynamics of the resource stock ($L_1$)

The rate of change in the resource stock, equivalent to the production rate, is described by the following differential equation:

$$\frac{dL_1}{dt} = -k_2 L_1 L_2 \tag{22}$$

This expression implies that production activity is positively influenced by both the remaining resource stock ($L_1$) and the capital available for oil production and utilization ($L_2$). In other words, the larger the reserve and the more developed the infrastructure, the higher the production rate. The evolution of the variations is addressed in Section 4.3 in the context of production.

EROIext is defined by the equation:

$$EROIext = \frac{\eta k_2}{k_3} L_1 \qquad (23)$$

Since $L_2$ and $\eta$ are constants or known values, the evolution of EROIext is directly linked to the value of $L_1$. Therefore, the temporal behavior of EROIext and the entropy ratio is fundamentally governed by the dynamics of $L_1$.

**4.2 Consideration of the relationship between capital stock ($L_2$) stock behavior and EROIext**

In this model, the $L_2$ stock is defined as the capital required for the production, refining, transportation, and utilization of crude oil. Figure 9 shows the temporal variation of capital. The differential equation representing the rate of change of the $L_2$ stock is:

$$\frac{dL_2}{dt} = \eta k_2 L_1 L_2 - k_3 L_2 \qquad (24)$$

The first term on the right-hand side is the production rate ($dL_1/dt$) multiplied by the coefficient $\eta$. The interaction between production and each stock will be discussed in Section 4.3. In the SCLV model, part of the production is supplied to the $L_2$ stock as capital for oil production. Therefore, production acts positively on the energy change of the $L_2$ stock.

The second term on the right-hand side represents the amount of low-entropy energy resources within the stock that are converted into high-entropy energy and discharged outside the system through the processes of energy production and consumption. As the scale of capital increases, the amount of low-entropy resources consumed also increases, so this term is proportional to the $L_2$ stock. Because it represents energy leaving the stock, it acts negatively on the energy change of the $L_2$ stock.

The variation of the $L_2$ stock is determined by the interaction between these two terms, and its temporal change is shown in Figure 9. Since $k_3$ is constant, the proportion of energy discharged outside the system to the total $L_2$ stock does not change. Therefore, the bell-shaped curve of the $L_2$ stock is considered to result from the decreasing amount of energy supplied relative to the energy consumed. In other words, the cause is believed to be the decline in EROIext, which represents the cost-effectiveness of oil (blue solid line in Figure 8).

Because $\eta$ is constant, the decrease over time in the energy supplied to the $L_2$ stock is due to the decline in production. The reduction in production caused by the interaction between $L_1$ and $L_2$ will be discussed in Section 4.3.

The relationship between the $L_2$ stock and EROIext also emerges when considering the conditions for the peak. The condition for the $L_2$ stock's energy to peak is when $dL_2/dt = 0$. Setting Eq. (24) to zero and rearranging yields:

$$\frac{\eta k_2}{k_3} L_1 = 1 \qquad (25)$$

The left-hand side is EROIext, meaning that the condition for the $L_2$ stock's energy to peak is EROIext=1. This has also been noted in the work of Perissi et al. (2021). In our model, EROIext falls below 1 in 2081, the year when the $L_2$ stock's energy reaches its peak. An EROIext below 1 means that the amount of oil-derived energy available for societal use is less than the energy required for oil production, implying that oil loses its societal value. It is particularly interesting that the timing of the loss of oil's societal value coincides with the point at which the capital needed for oil production and use begins to decline. This relationship, along with its connection to the production peak, will be examined in greater depth in Section 4.4.

Beyond the factors considered in this model, another possible reason for the decrease in the $L_2$ stock's rate of change is the increase in complexity within the stock resulting from the expansion of oil production capital. Tainter (1988) has explained, using the collapse of ancient empires as an example, that increased complexity in a system can influence its energy dynamics. According to Tainter (1988), as ancient empires expanded by conquering multiple regions, the complexity of the governance system increased accordingly. To manage this growing complexity, energy consumption increased, and when this consumption reached a critical point, the empire began to decline.

A similar phenomenon is likely to exist in oil production systems. For example, energy is consumed to adjust production volumes to stabilize market prices. As oil production systems expand and develop, they come to include both high- and low-productivity oil fields, and the differences in productivity between fields become more complex. This makes balancing oil production more difficult and increases the energy required for production adjustments.

One way to incorporate the effect of complexity into Eq. (25) is to add a term involving a coefficient multiplied by a power of $L_2$, which acts negatively on the rate of change of the $L_2$ stock (Eq. (26)). When such a complexity term is added, as the energy in $L_2$ increases, the energy consumption of the $L_2$ stock also rises, leading to an earlier point where $dL_2/dt = 0$. As a result, the energy peak of the $L_2$ stock is expected to occur earlier than in the present model.

$$\frac{dL_2}{dt} = \eta k_2 L_1 L_2 - k_3 L_2 - k_4 (L_2)^n \qquad (26)$$

## 4.3 Consideration of the relationship between production behavior and EROIext

Production is defined as $dL_1/dt$ and given by:

$$\text{Production} = k_2 L_1 L_2 \tag{27}$$

As stated in Section 4.1, when the remaining reserves are larger or the capital of the oil production system is greater, production becomes more active. Therefore, both $L_1$ and $L_2$ act positively on production.

Investigating the relationship between production (green solid line in Figure 6) and EROIext (blue solid line in Figure 8), it is observed that EROIext, which follows the same trend as $L_1$, declines more rapidly when production is increasing and more slowly when production is decreasing. Two possible underlying factors explain the relationship between production and EROIext.

The first is the progression toward higher-order recovery methods as oilfield development advances. When extracting crude oil from oilfields, there are several processes: primary recovery, which recovers spontaneously flowing crude oil; secondary recovery, which injects seawater or gas into the reservoir; and tertiary recovery, which alters the physical or chemical properties of rocks and reservoir fluids to enhance oil recovery. Recovery efficiency varies by method. According to U.S DoE (2020), primary recovery typically captures only ~10% of the oil in place, secondary methods such as water or gas injection can yield an additional 20–40%, and enhanced oil recovery (EOR, i.e., tertiary recovery) may ultimately recover 30–60% or more of the original oil in place. Higher-order recovery methods generally require greater costs and provide diminishing additional yields, which contributes to the decline in EROI as oilfield production advances.

The second factor is the principle of best-first. This well-established principle in economics and resource science states that humans tend to exploit the highest-quality resources first, followed by lower-quality resources. This tendency has been observed in agriculture and mining (e.g., Lovering, 1969; Murphy et al., 2011).

As oilfield development advances, the shift to higher-order recovery methods aligns with the economic "best-first principle" (Ricardo), whereby the most easily extractable and high-EROI resources are exploited first, followed by progressively costlier options, leading to lower overall EROI over time (Lambert et al., 2014). Over time, low-cost resources are depleted, and oil-dependent societies must develop higher-cost resources. However, oil extracted from such high-cost resources has a lower EROIext due to the greater input energy required.

In our model, the production curve is bell-shaped, with oil production peaking in 2042. Perissi et al. (2021) state that in the SCLV model, production peaks when the following

condition holds:

$$EROI = 1 + \frac{k_2}{k_3}L_2 \qquad (28)$$

Calculating the difference between EROIext and the right-hand side of Eq. (28) in our model shows that the smallest difference occurs in 2042. This confirms that, even when applying the SCLV model to oil, a relationship like that in Eq. (28) exists among the oil peak, EROIext, and oil production stocks.

Looking at Eq. (28), if the year of the oil peak is predetermined, a larger EROI appears to correspond to a larger $L_2$ stock. However, a larger EROI implies a more efficient energy production system, which would likely result in a larger $k_2$ (the parameter adjusting production) and a smaller $k_3$ (the parameter adjusting energy loss). Therefore, a larger EROI may increase the coefficient on the right-hand side of Eq. (28), meaning that a larger $L_2$ stock is not guaranteed. Nevertheless, when $k_2$ is large and $k_3$ is small, the model suggests that energy accumulates more easily in the capital stock, so it is highly likely that a larger EROI would correspond to a larger $L_2$ stock.

The peak in oil production (oil peak) has been the subject of extensive debate. It was first proposed by Hubbert (1956), who applied the logistic equation—a mathematical model of population dynamics—to oil production trends. Hubbert argued that oil production follows a symmetrical bell-shaped curve, the Hubbert curve, and that the oil peak occurs when roughly half of the reserves have been consumed. In 1956, Hubbert predicted that the oil peak would occur around the year 2000. Later, Campbell and Laherrère (1998) re-examined the Hubbert curve and updated the reserve estimates, predicting the oil peak to occur between 2004 and 2005. Furthermore, Sorrell et al (2010) analyzed over 500 studies on oil depletion and suggested that the oil peak was highly likely to occur by 2030. According to Laherrère et al (2022), conventional oil has already peaked in 2019, while all liquids are estimated to peak around 2040. These peaks are expected to have significant economic, political, and sustainability impacts.

In our SCLV model, the cumulative oil production up to the peak year of 2042 was approximately 37.5% of the ultimately recoverable reserves, differing from the conventional view that the oil peak occurs when about half the reserves have been consumed. When considering the oil peak within the framework of the SCLV model, it becomes possible to explore relationships between the oil peak, the oil production system, and EROI that are difficult to address in conventional studies. In this respect, our study significantly differs from existing oil-related research.

**4.4 Relationship between the lag in capital peak and production peak and its societal**

**implications**

Investigating the trajectories of capital stock (orange solid line in Figure 6) and oil production (green solid line in Figure 6), we can observe that the capital peak occurs in 2081, whereas the oil production peak (oil peak) occurs in 2041. Thus, the capital peak lags 40 years behind the production peak. Notably, the timing of the capital peak coincides with the point at which EROIext falls to 1, marking the moment when oil loses its societal value. This implies that oil loses its societal value 40 years after production has peaked.

The occurrence of the capital peak after the oil peak indicates that there is a period during which capital for utilizing oil continues to expand, despite a decline in oil production. Such investment in exploiting a declining resource can be considered growth without substance—capital stock in the oil production sector grows, but lacks long-term viability. Similarly, a society that continues to develop while depending on a declining resource experiences illusory growth. This illusion becomes evident when the expansion of oil production capital is no longer sustainable—that is, when oil loses its societal value. In the context of the society assumed by this model, the critical point for degradation described by Tainter (1988) corresponds to the moment when EROIext falls below 1, after which both oil production capital and society as a whole are expected to decline.

The main cause of the peak lag is that oil demand increases—or at least does not decrease—even as oil production falls or its growth rate slows. In the present model, alternative energy sources are not considered, and the society is assumed to be highly dependent on oil. Because oil retains high importance in such a society, capital investment in oil development continues even if production decreases, creating the observed lag between the oil peak and the capital peak.

In reality, progress in energy mix diversification may reduce the necessity for oil production compared with the assumptions of this model, thereby potentially shortening the lag. However, OPEC Secretariat (2022) projects that global oil demand will continue to increase between 2021 and 2045. Therefore, in actual society as well, a certain period of lag between the oil peak and the peak in capital for oil production is likely to occur. This means that even after oil production peaks and begins to decline, the scale of capital dedicated to oil production and utilization may continue to expand, resulting in growth that is essentially unsustainable.

Conventional oil peak studies have not examined the relationship between the oil peak and the peak in capital for oil production and utilization. The ability to explore this relationship is a key contribution of the present study to the broader field of resource peak research.

**4.5 Comparison with previous studies**

Research on EROI has been active, but most studies focus on a single year, calculating the ratio of total energy output to total energy input over the life cycle (Murphy et al., 2022). For example, Brockway et al. (2019) used life-cycle analysis to estimate the EROIpou of fossil fuels from 1995 to 2011. EROIpou is defined as the ratio of the net energy delivered to end users to the total energy expended throughout the entire supply chain, including extraction, processing, transportation, and distribution. The life-cycle input–output approach has been studied extensively over many years, and its quantitative accuracy in estimating EROI is therefore relatively well established. However, because it relies on historical energy input–output data, it is difficult to project future trends in EROI, and such studies are typically limited to analyzing EROI itself.

In contrast, the present study analyzes EROIext using the SCLV model. While the method of defining stocks and the results of sensitivity analysis indicate that guaranteeing strict quantitative accuracy is challenging, the ability to represent the qualitative behavior of EROIext up to the year 2100 is a feature not found in previous EROI research. Furthermore, the SCLV model allows analysis of the relationships between EROI and other parameters such as oil production capital and production volume, enabling consideration of EROI within the broader interaction between production capital and resources—an aspect absent from many EROI studies. By incorporating the concept of entropy, the model also offers the potential to address impacts that are difficult to analyze solely from an energy-based perspective. Naturally, there remain numerous factors that cannot be captured in a simple two-stock model; these will be discussed in Section 4.7.

**4.6 Entropy Ratio and Systemic Implications**

The entropy ratio, $\Delta S/\Delta S_1$, represents the expected reduction in entropy when energy is supplied to a system. In this study, it is expressed using the SCLV model parameters as follows:

$$\frac{\Delta S}{\Delta S_1} = 1 - \frac{\eta k_2}{k_3} L_1 \qquad (29)$$

As noted in the methodology, the entropy ratio can be regarded as a measure of a system's efficiency in reducing entropy, sharing similar properties with EROIext, which measures efficiency from an energy perspective. However, because EROIext focuses solely on energy, it is difficult to incorporate environmental factors such as $CO_2$ emissions or water pollution. In contrast, entropy can encompass these effects through the concept of environmental entropy increase. Therefore, the entropy ratio can be viewed as a comprehensive indicator of the broader societal impact of an energy production system.

In terms of its influence on the model, the entropy ratio behaves similarly to EROIext, as the second term on the right-hand side of Eq. (29) corresponds to EROIext. The modeled behavior of the entropy ratio suggests that the entropy-reducing effect of oil diminishes over time and eventually disappears, at which point energy input into the system leads to an overall increase in entropy. In the present model, this turning point occurred in 2081. The reduction in entropy-reduction capacity is likely due to the increasing entropy released to the external environment ($S_2$ in Figure 2) during oil production, driven by the same factors that cause EROIext to decline—such as increased energy consumption, more complex production processes, and environmental impacts like $CO_2$ emissions and water contamination.

While the model accounts only for entropy generated during oil production and the construction of oil-use infrastructure, in reality, increasing system complexity itself may further accelerate entropy generation. If such complexity effects are included, the expansion of capital would likely result in an even faster increase in entropy, thereby reducing the oil production system's ability to decrease entropy.

### 4.7 Limitations and Future Directions

In constructing the model and integrating the studies of Hu et al. (2021) and Perissi et al. (2021), several assumptions were made. This section outlines these assumptions and limitations, followed by a discussion of the study's future potential.

One key assumption was made when incorporating the energy required for transportation and use infrastructure into capital, using results from Hall et al. (2009). Hall et al. (2009) assumed that the U.S. economy operated entirely on domestic oil and that energy transport occurred exclusively via automobiles. Consequently, the present study does not account for the influence of non-oil fossil fuels, nor does it consider alternative oil transportation modes such as shipping in the capital stock. Therefore, the results should be interpreted with caution, acknowledging the exclusion of factors such as energy mix diversification and transport mode variety. Moreover, applying a U.S.-specific study to a model intended to represent the global oil production system also introduces limitations.

Particular attention should be paid to the role of the energy mix. The model assumes a society powered solely by oil. However, according to Ritchie et al. (2022), the share of oil in global primary energy consumption between 1965 and 2010 ranged from approximately 35–50%. During this period, oil production was already influenced by the energy mix. Therefore, the oil production data used as the model baseline differ in nature from the hypothetical oil-only society assumed in the model. If a fossil fuel energy mix were considered, oil dependence would be lower than in the current model, likely leading to slower growth in capital stock and production volume, slower depletion of resources, a more gradual decline in both EROI and entropy ratio, and a later point at which oil loses its societal value.

The potential role of renewable energy in the energy mix must also be considered. Some studies suggest that the EROI of renewable energy may exceed that of fossil fuels, implying that increased adoption could stabilize global EROI (Brockway et al., 2019). However, current renewable energy's share of global primary energy is only about 14%, and expanding this share requires infrastructure built using fossil fuel-derived energy. Therefore, oil production is likely to continue increasing until renewable energy capacity is sufficiently expanded, with corresponding capital stock growth. Once renewables reach a substantial share, production and capital growth may stabilize; however, if production has already begun to decline or capital has peaked before this point, net energy supply to society could be disrupted, potentially leading to societal decline. The timing of renewable energy expansion will thus influence the dynamics of all stocks and production volume. By extending the SCLV model to qualitatively capture the relationship between fossil fuel resources, development capital, production volume, and renewable energy share, the societal impacts of renewable energy adoption can be examined from an energy perspective. This could help design renewable energy introduction plans that avoid societal instability, making the SCLV model a potentially valuable tool for energy policy.

When expressing the entropy ratio in terms of the SCLV model parameters, this study assumed that the EROI in Hu et al. (2021) corresponds to EROIext. To evaluate this assumption's validity, it is important to clarify what is included in the "Energy resource exploitation system" in Figure 2. In Hu et al. (2021), this system is interpreted as "a system that, when supplied with low-entropy resources from the socioeconomic system, extracts energy from resources and delivers it to the socioeconomic system." This definition implies that the "Energy resource exploitation system" includes not only the extraction of energy from resources but also the energy and infrastructure needed for transportation and delivery. Therefore, it is reasonable to interpret the EROI in Hu et al. (2021) as EROIext.

A comparison of the entropy-based energy production system model in Figure 2 and the SCLV model reveals both similarities and differences. Both systems consume low-entropy resources and dissipate high-entropy outputs to the environment. The energy production capital stock in the SCLV model can thus be seen as having a dissipative structure that maintains internal entropy levels by releasing high-entropy outputs. A key difference, however, is that the SCLV model does not account for the energy of the "Social economic system". In Figure 2, energy investment into the production system comes from the socioeconomic system, whereas in the SCLV model, it comes directly from the resource stock. Furthermore, while Figure 2 includes the consumption of energy within the socioeconomic system, the SCLV model treats all non-oil capital energy destinations as external to the system. This means the current SCLV model cannot account for entropy changes resulting from societal energy consumption. Since, in reality, decisions about energy investment into the production system are made by the socioeconomic system rather than the production system itself, Figure 2 arguably reflects real-world dynamics more accurately.

To incorporate energy and entropy dissipation into the SCLV model as in Figure 2, a third stock—representing the socioeconomic system—could be added. The inclusion of a socioeconomic stock is also noted in Perissi et al. (2021). A combined three-stock model integrating the SCLV framework with entropy dissipation could resemble the structure shown in Figure 10. In such a model, energy from the resource stock would first move to the socioeconomic stock, and from there into the energy production system. When energy is consumed in either the socioeconomic or production systems, high-entropy matter would be released to the environment. If such a model were formulated mathematically for both energy and entropy and applied to a specific resource, it could enable a more realistic analysis of the relationship between energy and entropy in society.

In addition, the current dynamics of the global energy transition must be acknowledged. Although renewable energy has the potential to eventually achieve higher EROI values than fossil fuels (Brockway et al., 2019), its present deployment faces significant challenges, including intermittency, grid integration constraints, storage limitations, and uneven investment across regions (Krishnan et al., 2024). These barriers imply that renewable expansion may not progress smoothly enough to offset declining oil production in the near term (Heard et al., 2017). Consequently, societies will remain reliant on oil for several decades, necessitating continued—though carefully calibrated—investment in oil development. Such investment is essential to maintain short- to medium-term energy security and economic stability (Cherp et al., 2017; Jewell et al., 2018; IEA, 2022). However, beyond a certain point, sustained investment in oil risks creating stranded assets, misallocating capital, and delaying the transition toward sustainable systems. Thus, oil investment embodies a dual nature: it is indispensable in the short run but increasingly unsustainable in the long run. Recognizing and managing this duality is crucial for aligning energy policy with both near-term stability and long-term resilience.

## 5. Conclusion and Policy Implication

This study applied a system dynamics model to investigate the long-term dynamics of global oil production, incorporating both physical resource limits and thermodynamic efficiency. Two central findings emerge. First, global oil production is projected to peak around 2041, after which output declines irreversibly. Second, the quality of oil as an energy source—measured by the extended Energy Return on Investment (EROIext)—falls below unity by 2081. At this point, oil ceases to provide net energy to society, regardless of geological reserves, and becomes a drain on resources. The model also reveals a forty-year lag between the production peak and the peak of capital devoted to oil exploitation, underscoring the risk of sustained investment in a system that no longer delivers sustainable returns.

These results highlight both the risks and necessities of investment in oil. On the one hand, oil will remain a central component of the global energy mix for several decades. A premature halt in investment could exacerbate supply shortages, heighten price volatility, and increase geopolitical instability. Continued, targeted investment in oil development is therefore essential in the near and medium term to ensure energy security and provide stability during the transition. On the other hand, the persistent decline in EROIext and the eventual loss of oil's net energy contribution demonstrate the limits of such investment. Over-investment risks creating stranded assets, misallocating capital that could be used for renewable energy development, and delaying the inevitable shift to more sustainable systems.

Importantly, these findings must be interpreted against the current reality that energy transitions to renewable sources are not proceeding smoothly. Despite significant progress in deployment, challenges remain in scaling renewable energy to a level that can replace fossil fuels, including intermittency, storage limitations, grid integration, and uneven investment across regions. This means that in the foreseeable future, societies will continue to rely on oil and other fossil fuels as part of the energy mix. Therefore, policy must not only aim for rapid expansion of renewables but also acknowledge the transitional role of oil and design strategies that manage both sides of this dual reality.

The implications for policy are fourfold:

1. **Integrate net-energy and thermodynamic indicators into energy planning.** Policymakers should assess projects not only by economic costs or reserves but also by their capacity to provide net energy. This would prevent overvaluation of low-quality resources and ensure that transition planning reflects both physical and thermodynamic realities.

2. **Maintain short-term investment for energy security.** Given the uneven pace of renewable deployment, calibrated oil investment remains necessary to avoid supply shocks and to sustain economic activity. Such investments should be transparent, limited in scope, and embedded within broader transition strategies.

3. **Address the limitations of current renewable transitions.** Recognizing that renewables face integration and scalability challenges, governments must invest in complementary measures such as grid modernization, storage technologies, and demand-side management. This will reduce systemic vulnerability while renewables expand.

4. **Redirect long-term capital toward sustainable systems.** Even as short-term oil investments continue, capital must increasingly be reallocated toward renewables and efficiency improvements to prevent lock-in and stranded assets. Coordinated policies that gradually shift incentives will help balance stability today with resilience tomorrow.

In conclusion, the challenge is not only the finite volume of oil but also its declining thermodynamic quality, set against the imperfect and uneven progress of renewable transitions. Policies that continue to prioritize oil without recognizing its limits risk undermining long-term resilience, while policies that assume a frictionless transition may underestimate near-term risks. A balanced approach—acknowledging the current difficulties of renewable deployment while preparing for oil's eventual decline—offers the most realistic pathway for sustaining energy security and guiding societies through the transition to a more resilient energy future.

**Acknowledgements**
We are grateful to Vincent Court and François Fizaine for kindly providing the dataset they calculated in their study (Court and Fizaine, 2017), which was used in the preparation of Figure 3 in this paper.

**Declaration of generative AI and AI-assisted technologies in the writing process**
During the preparation of this work, the author (Jun Matsushima) used ChatGPT (OpenAI, San Francisco, CA, USA) in order to improve the clarity of English expressions and to check grammar. After using this tool, the author (Jun Matsushima) reviewed and edited the content as needed and take full responsibility for the content of the published article.

**References**

Bardi, U., Lavacchi, A., 2009. A simple interpretation of Hubbert's model of resource exploitation. Energies 2, 646–661. https://doi.org/10.3390/en20300646.

BP, 2002. BP statistical review of world energy 2002. BP, London.

BP, 2010. BP statistical review of world energy 2010. BP, London.

BP, 2021. BP statistical review of world energy 2021. BP, London.

Brand-Correa, L.I., Brockway, P., Copeland, C.L., Foxon, T.J., Owen, A., Taylor, P.G., 2017. Developing an input–output based method to estimate a national-level energy return on investment (EROI). Energies 10, 534. https://doi.org/10.3390/en10040534.

Brockway, P.E., Owen, A., Brand-Correa, L.I., Hardt, L., 2019. Estimation of global final-stage energy-return-on-investment for fossil fuels with comparison to renewable energy sources. Nat. Energy 4, 612–621. https://doi.org/10.1038/s41560-019-0425-0.

Campbell, C.J., Laherrère, J.H., 1998. The end of cheap oil. Sci. Am. 278, 78–83.

Cherp, A., Vinichenko, V., Jewell, J., Suzuki, M., Antal, M., 2017. Comparing electricity transitions: A historical analysis of nuclear, wind and solar power in Germany and Japan. Energy Policy 101, 612–628.

Cleveland, C.J., Costanza, R., Hall, C.A.S., Kaufmann, R.K., 1984. Energy and the U.S. economy: A biophysical perspective. Science 225, 890–897. https://doi.org/10.1126/science.225.4665.890.

Court, V., Fizaine, F., 2017. Long-term estimates of the energy-return-on-investment (EROI) of coal, oil, and gas global productions. Ecol. Econ. 138, 145–159. https://doi.org/10.1016/j.ecolecon.2017.03.002.

Gagnon, N., Hall, C.A.S., Brinker, L., 2009. A preliminary investigation of energy return on energy investment for global oil and gas production. Energies 2, 490–503. https://doi.org/10.3390/en20300490.

Hall, C.A.S., 1988. An assessment of several of the historically most influential theoretical models used in ecology and of the data provided in their support. Ecol. Modell. 43, 5–31. https://doi.org/10.1016/0304-3800(88)90059-9.

Hall, C.A.S., Balogh, S., Murphy, D.J.R., 2009. What is the minimum EROI that a sustainable society must have? Energies 2, 25–47. https://doi.org/10.3390/en20100025.

Hall, C.A.S., Cleveland, C.J., Kaufmann, R.K., 1986. The ecology of the economic process: Energy and resource quality. Energy Syst. Policy 9, 602–625.

Heard, B.P., Brook, B.W., Wigley, T.M.L., Bradshaw, C.J.A., 2017. Burden of proof: A comprehensive review of the feasibility of 100% renewable-electricity systems. Renew. Sustain. Energy Rev. 76, 1122–1133.

Hu, Y., Feng, L., Hall, C.A.S., Tian, D., Zhao, L., 2011. Analysis of the energy return on investment (EROI) of the huge Daqing oil field in China. Sustainability 3, 2323–2338. https://doi.org/10.3390/su3122323.

Hu, Y., Hall, C.A.S., Wang, J., Feng, L., Poisson, A., Wei, W., 2021. An explanation of energy return on investment from an entropy perspective. Front. Energy Res. 9, 633528. https://doi.org/10.3389/fenrg.2021.633528.

Hubbert, M.K., 1956. Nuclear energy and the fossil fuels. Shell Development Company, Houston.

IEA, 2022. World Energy Outlook 2022. International Energy Agency, Paris.

Japan Petroleum Development Association, 2018. Resource Assessment Study 2017 (Evaluation of Global Petroleum and Natural Gas Resources as of the End of 2015). JPDA, Tokyo.

Jewell, J., McCollum, D., Emmerling, J., Bertram, C., Gernaat, D.E., Krey, V., Paroussos, L., Berger, L., Fragkiadakis, K., Keppo, I., Saadi, N., 2018. Limited emission reductions from fuel subsidy removal except in energy-exporting regions. Nature 554, 229–233.

King, C.W., 2015. Comparing world economic and net energy metrics, Part 3: Macroeconomic historical and future perspectives. Energies 8, 12997–13020. https://doi.org/10.3390/en81212330.

Krishnan, M., Bradley, C., Tai, H., Devesa, T., Smit, S., Pacthod, D., 2024. The Hard Stuff: Navigating the physical realities of the energy transition. McKinsey Global Institute, New York.

Laherrère, J., Hall, C.A.S., Bentley, R., 2022. How much oil remains for the world to produce? Comparing assessment methods, and separating fact from fiction. Curr. Res. Environ. Sustain. 4, 100174.

Lambert, J.G., Hall, C.A.S., Balogh, S., Gupta, A., Arnold, M., 2014. Energy, EROI and quality of life. Energy Policy 64, 153–167.

Lotka, A.J., 1925. Elements of physical biology. Williams & Wilkins, Baltimore.

Lovering, T.S., 1969. Mineral resources from the land, in: Resources and Man. National Academy of Sciences, Washington, D.C., pp. 109–134.

Murphy, D.J., Hall, C.A.S., Powers, B., 2011. New perspectives on the energy return on (energy) investment (EROI) of corn ethanol. Environ. Dev. Sustain. 13, 179–202.

Murphy, D.J., Raugei, M., Carbajales-Dale, M., Rubio Estrada, B., 2022. Energy return on investment of major energy carriers: Review and harmonization. Sustainability 14, 7098.

Odum, H.T., 1973. Energy, ecology, and economics. Ambio 2, 220–227.

OPEC Secretariat, 2022. World Oil Outlook 2045. Organization of the Petroleum Exporting Countries (OPEC), Vienna.

Perissi, I., Bardi, U., Lavacchi, A., 2017. Dynamic patterns of overexploitation in fisheries. Ecol. Modell. 359, 285–292. https://doi.org/10.1016/j.ecolmodel.2017.06.013.

Perissi, I., Lavacchi, A., Bardi, U., 2021. The role of energy return on energy invested (EROEI) in complex adaptive systems. Energies 14, 8411.

Prigogine, I., Lefever, R., 1973. Theory of dissipative structures, in: Synergetics: Cooperative Phenomena in Multi-Component Systems. Vieweg+Teubner, Wiesbaden, pp. 124–135.

Ritchie, H., Roser, M., Rosado, P., 2022. Energy. Our World in Data. https://ourworldindata.org/energy (accessed 16 August 2025).

Smith, V.L., 1968. Economics of production from natural resources. Am. Econ. Rev. 58, 409–431.

Sorrell, S., Speirs, J., Bentley, R., Brandt, A., Miller, R., 2010. Global oil depletion: A review of the evidence. Energy Policy 38, 5290–5295.

Tainter, J.A., 1988. The collapse of complex societies. Cambridge Univ. Press, Cambridge. https://doi.org/10.1017/CBO9780511570985.

U.S. Department of Energy, 2020. Enhanced Oil Recovery. https://www.energy.gov/fecm/enhanced-oil-recovery (accessed 16 August 2025).

Volterra, V., 1927. Fluctuations in the abundance of a species considered mathematically. Nature 119, 12–13.

Wang, M., 2001. The ecological significance of the theory of dissipative structure. Soc. Sci. J. Colleges Shanxi 13, 23–24.

Wang, R., Ouyang, Z., 2012. Social-Economic-natural complex ecosystem and sustainability. Bull. Chin. Acad. Sci. 27, 337–345.

Xu, D., Wang, Z., Guo, L., 2004. Entropy analyses and distinguishing of industrial ecological system evolution based on dissipative structure theory. Manag. Sci. China 17, 51–56.
28

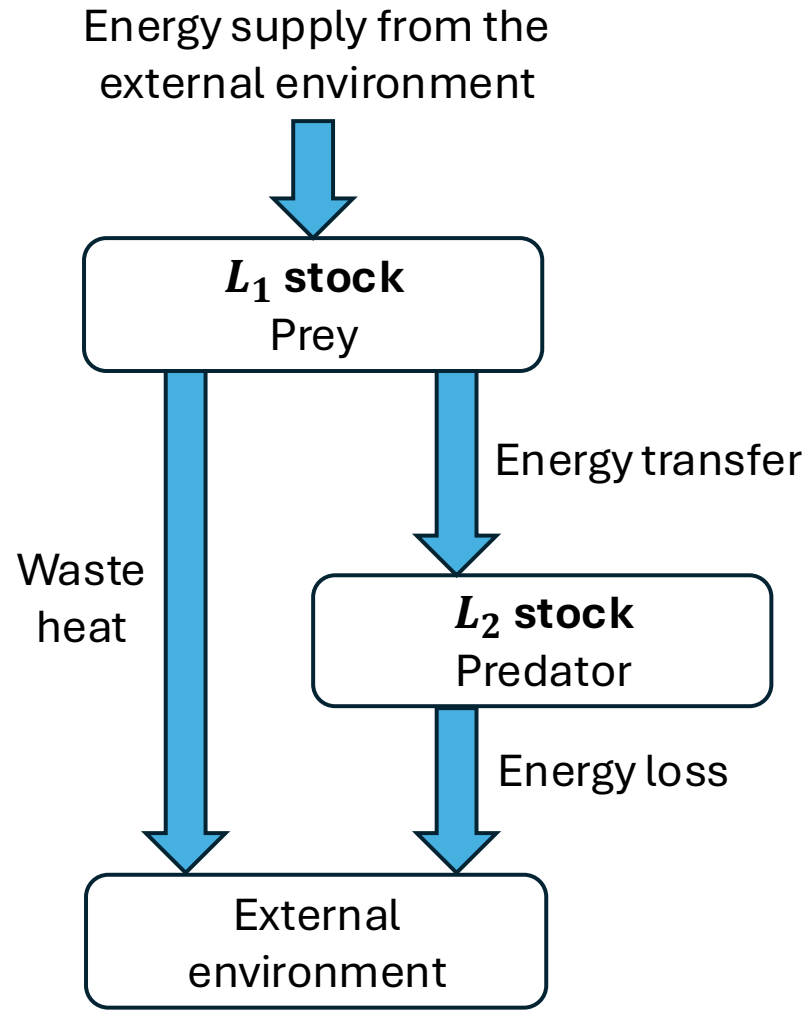

Figure 1. Energy flow diagram of the Lotka–Volterra (LV) model, showing the transfer of energy between prey and predator stocks, with dissipation to the external environment.

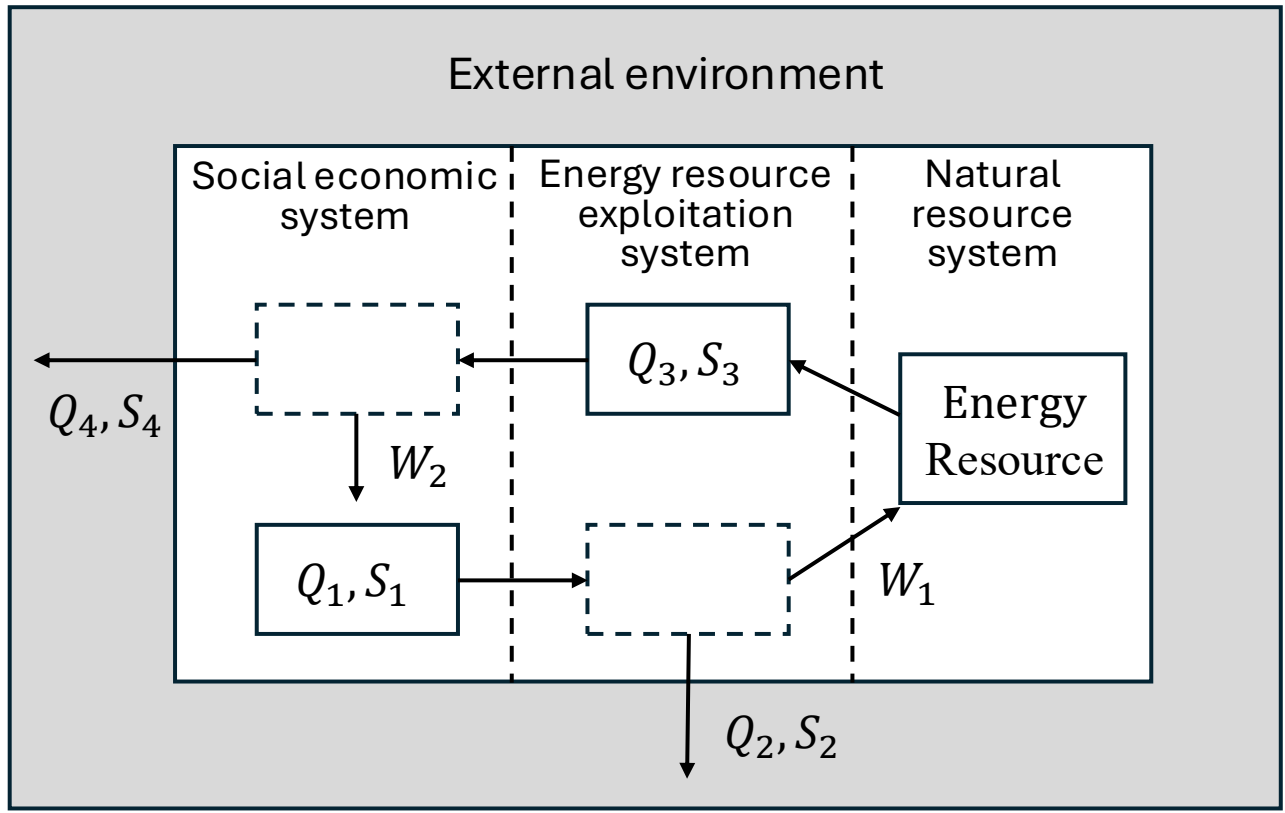

Figure 2. Analytical diagram of an energy production system based on a Carnot heat engine, illustrating energy and entropy flows among natural resource system, the energy resource exploitation system, and the social economic system (modified after [11]).

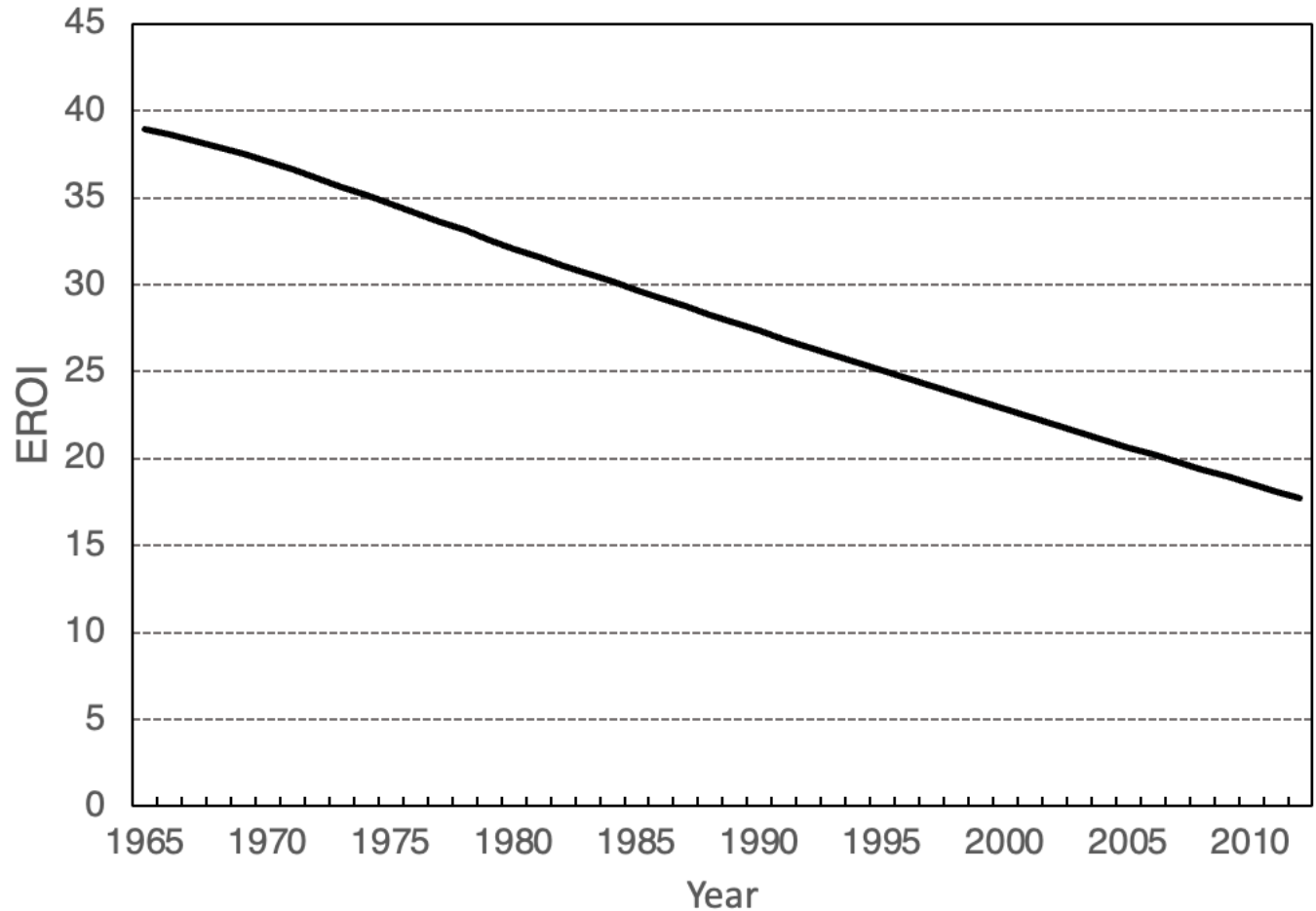

Figure 3. Historical trajectory of the energy return on investment (EROI) for global oil production. The data used in this figure were calculated by [23] and provided directly by the authors.

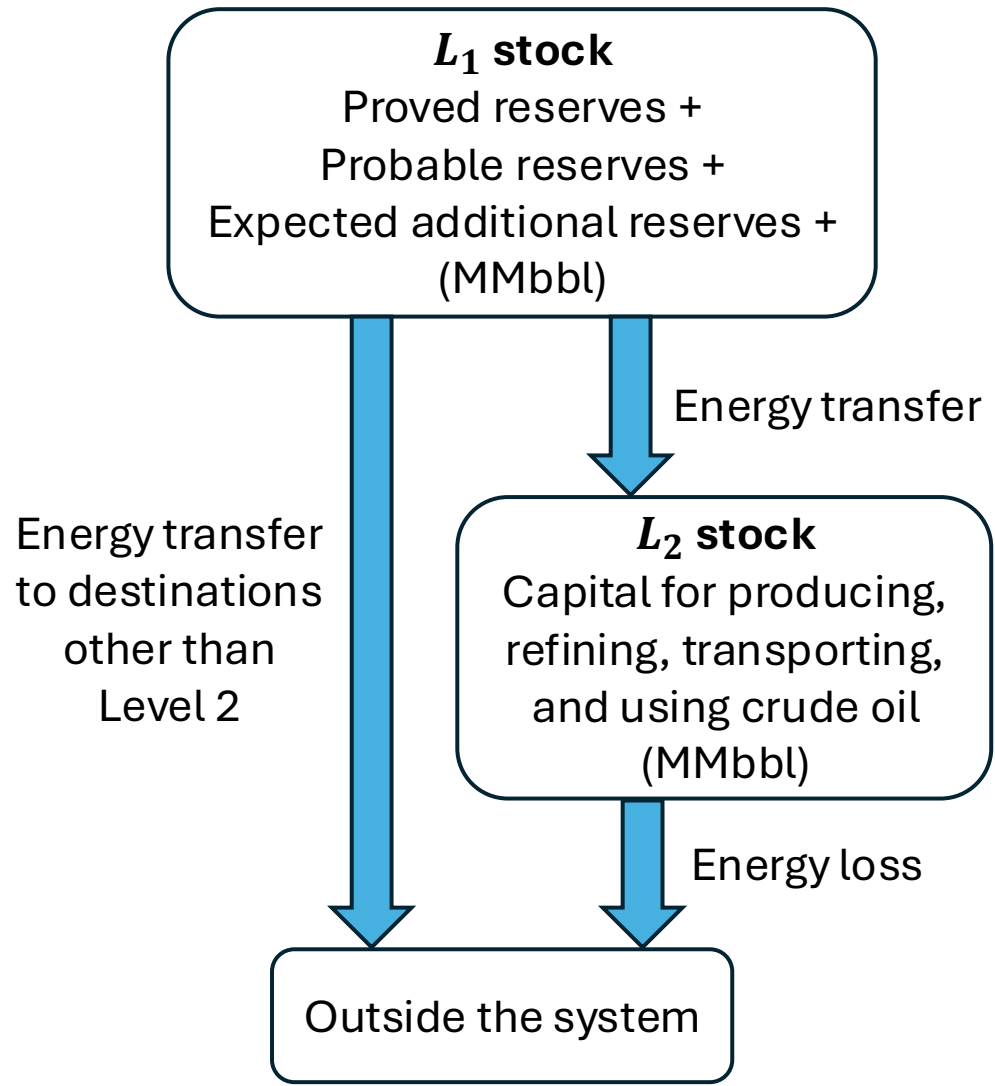

Figure 4. Energy flow diagram of the SCLV model developed in this study, representing energy transfer from the resource stock to the capital stock, and energy losses to the external environment as waste heat and non-usable energy.

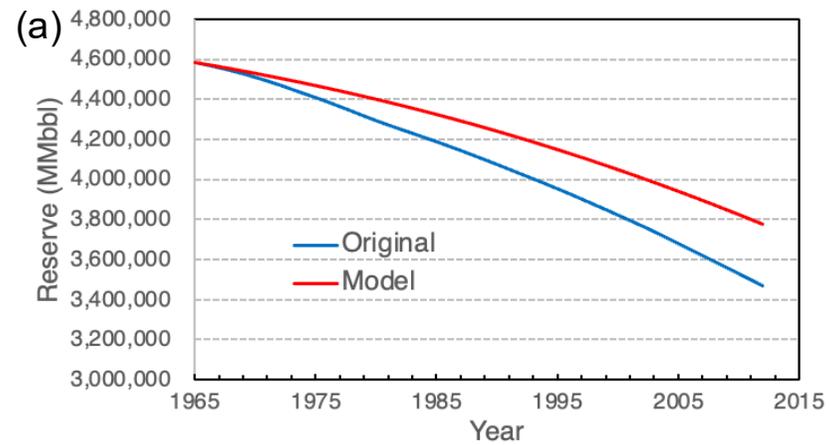
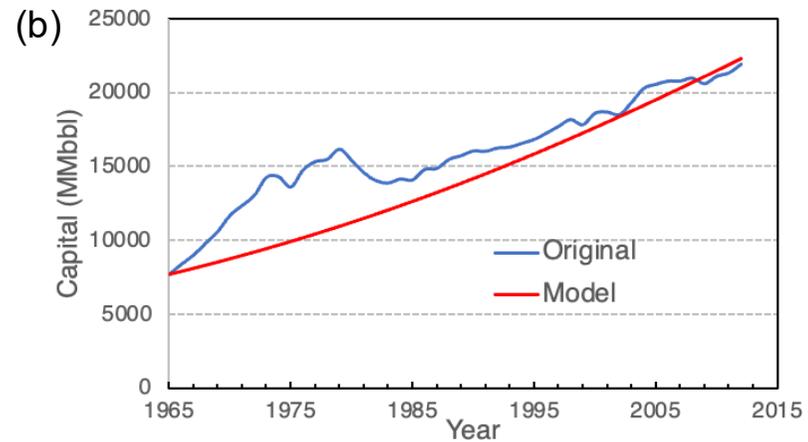
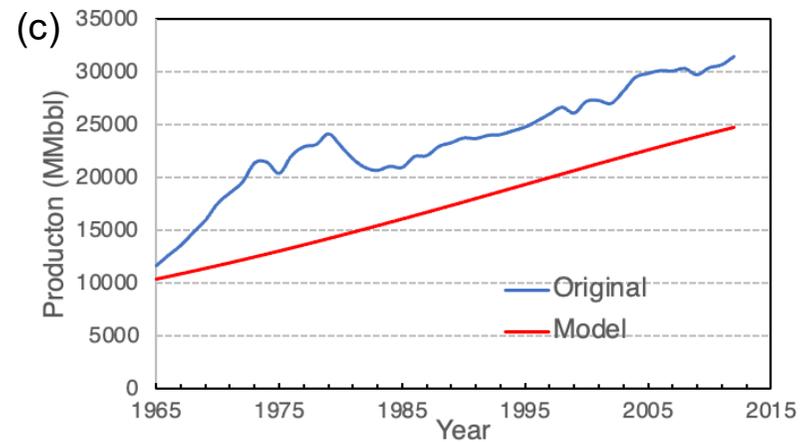

Figure 5. Comparison of baseline data and model results (1965–2012): (a) Resource stock (Reserve), (b) Capital stock, (c) Oil production. Blue lines denote original data, and red lines indicate modeled values.

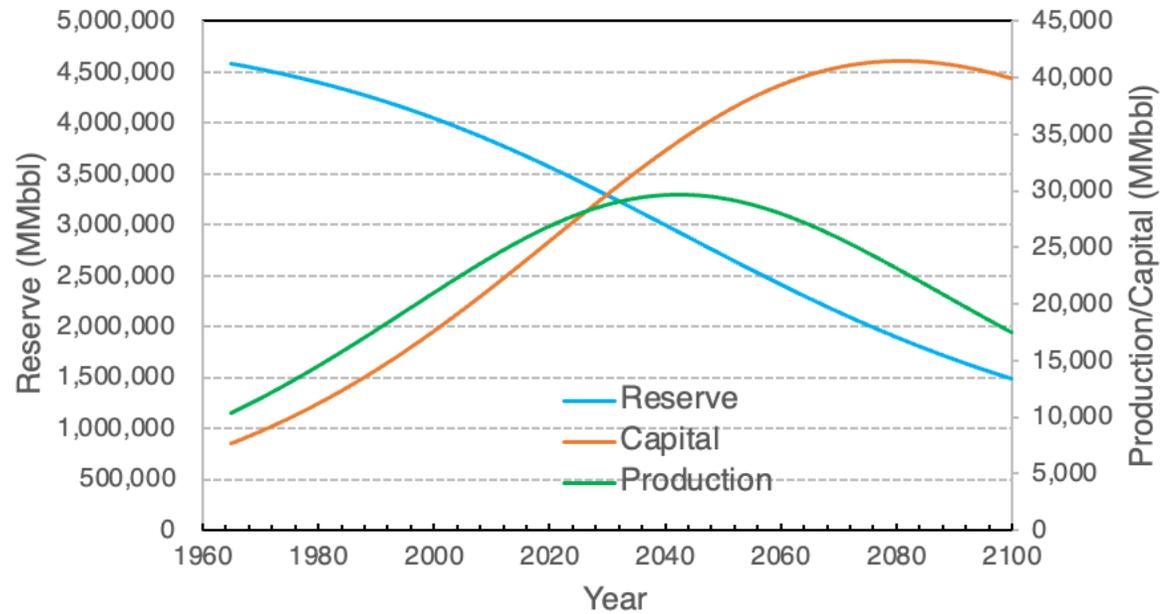

Figure 6. Projected dynamics of the resource stock (Reserve), capital stock, and oil production through 2100, showing peaks in oil production (2041) and capital stock (2081).

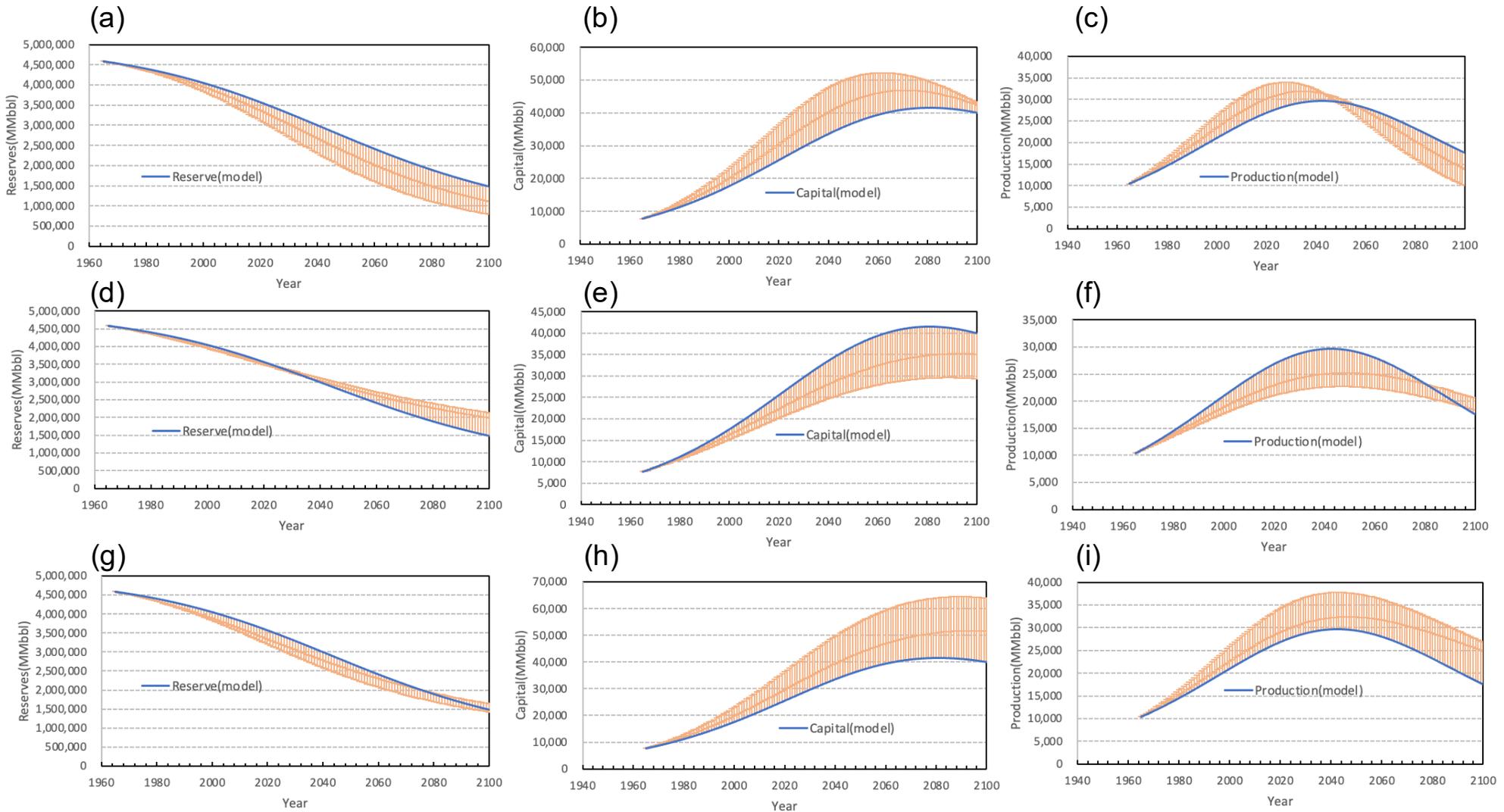

Figure 7. Sensitivity analysis of model behavior under parameter variations. Panels (a), (d), and (g) show the trajectories of the resource stock when varying $k_2$, $k_3$, and $\eta$, respectively. Panels (b), (e), and (h) present the trajectories of the capital stock under the same parameter variations. Panels (c), (f), and (i) illustrate the corresponding trajectories of oil production.

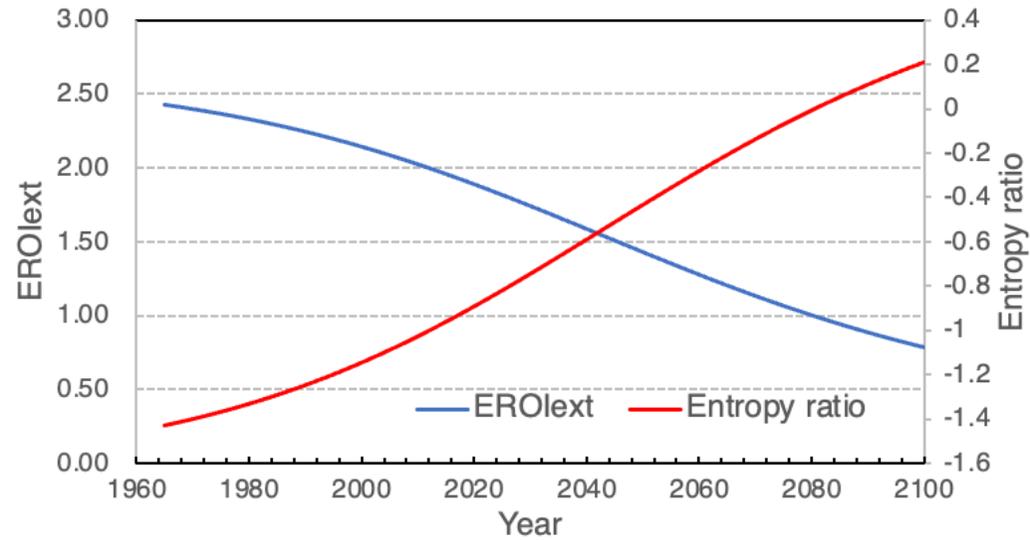

Figure 8. Trajectories of extended energy return on investment (EROIext, blue line) and entropy ratio ($\Delta S/\Delta S_1$, red line) from 1965 to 2100, showing the decline of EROIext below unity by 2081 and the steady increase of the entropy ratio.

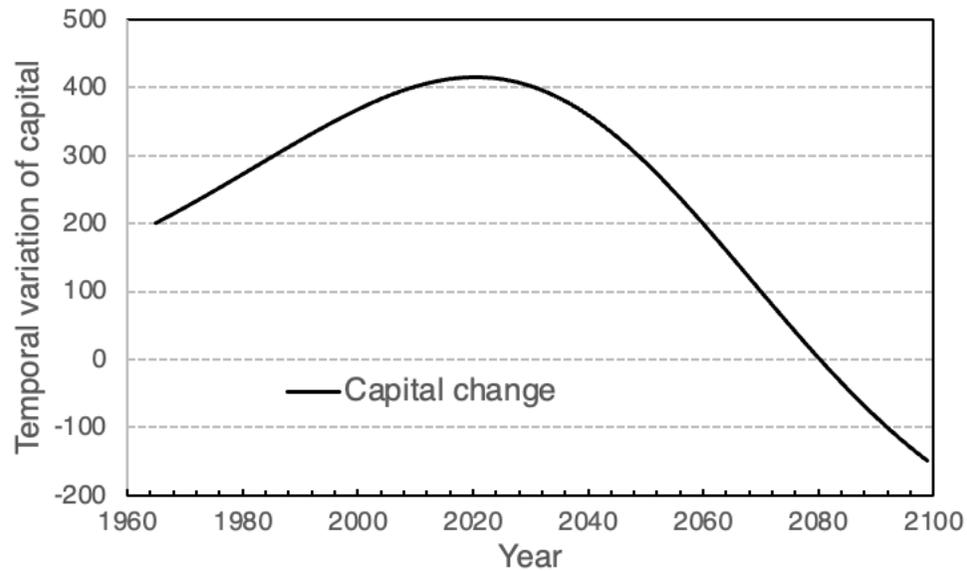

Figure 9. Temporal variation of the capital stock for petroleum production, refining, transportation, and utilization, illustrating its bell-shaped trajectory caused by declining EROIext.

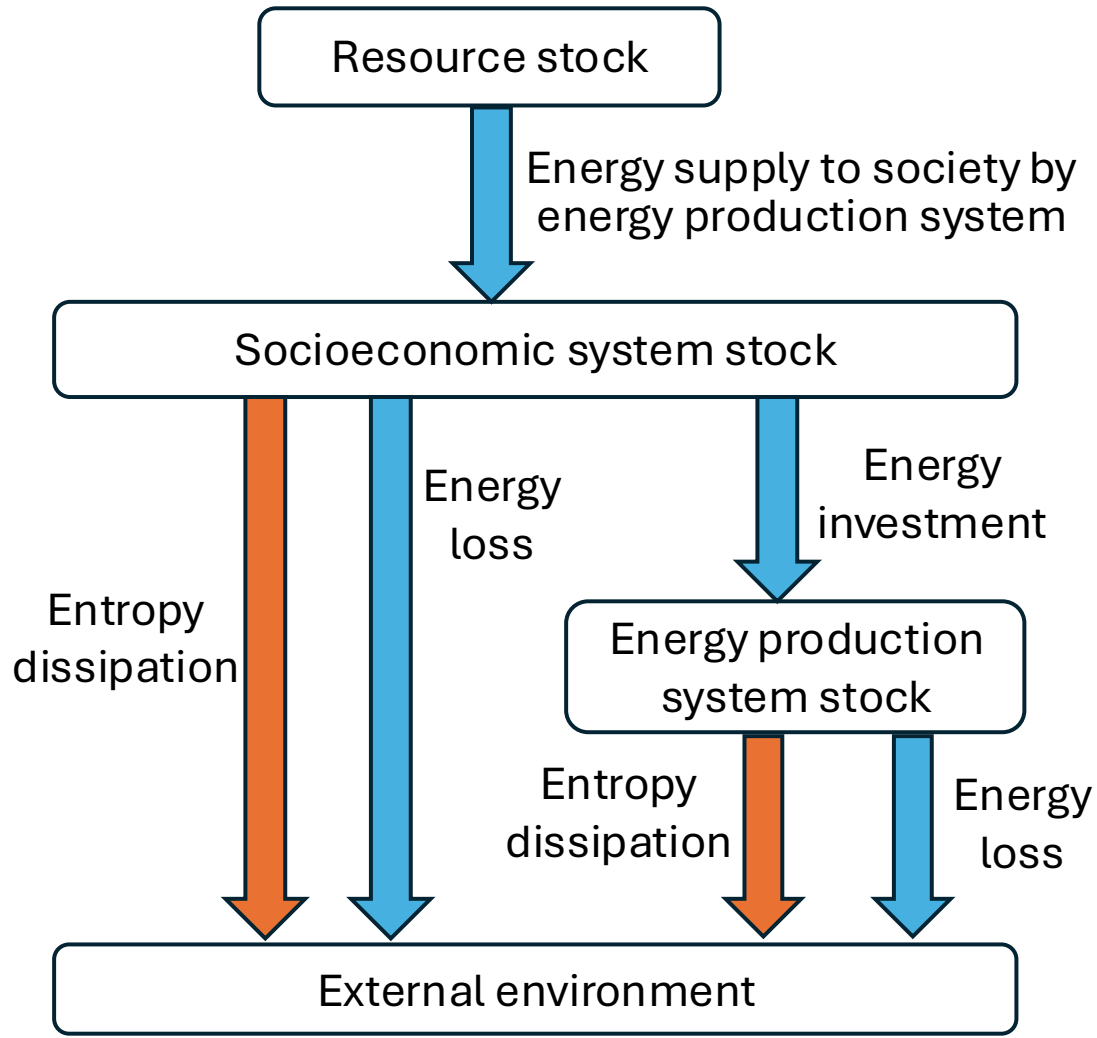

Figure 10. Extended SCLV model integrating a socioeconomic stock to incorporate energy and entropy dissipation, as discussed in the text. Energy flows sequentially from the resource stock to the socioeconomic stock and then to the production system, with high-entropy matter released to the environment. Adapted from the framework noted by [13].

Table 1. Fitting results of the SCLV model using baseline data. The table reports the estimated parameter values—including $k_2$, $k_3$, and $\eta$—together with the sum of squared error, which serve as the basis for the subsequent sensitivity analyses

| $k_2$ | $k_3$ | $\eta$ | Sum of Squared Errors |
|---|---|---|---|
| $2.9397 \times 10^{-7}$ | $1.8271 \times 10^{-2}$ | $3.2883 \times 10^{-2}$ | 26.1245 |

Table 2. Sensitivity analysis of the peak year of oil production under different parameter settings. The table compares the baseline SCLV model with cases where the parameters $k_2$, $k_3$, and $\eta$ are varied by $\pm 10\%$. The results indicate the corresponding shifts in the timing of the oil production peak relative to the baseline model.

|  | Model | $k_2 \times 1.1$ | $k_2 \times 9.0$ | $k_3 \times 1.1$ | $k_3 \times 0.9$ | $\eta \times 1.1$ | $\eta \times 0.9$ |
|---|---|---|---|---|---|---|---|
| Year of Peak | 2042 | 2035 | 2028 | 2049 | 2047 | 2047 | 2043 |
| Difference from Model |  | 7 years | 14 years | 7 years | 5 years | 5 years | 1 year |

Table 3. Sensitivity analysis of the peak year of capital stock under different parameter settings. The baseline SCLV model is compared with cases where the parameters $k_2$, $k_3$, and $\eta$ are varied by $\pm 10\%$. The results show how these parameter changes shift the timing of the capital stock peak relative to the baseline model.

|  | Model | $k_2 \times 1.1$ | $k_2$ 9.0× | $k_3 \times 1.1$ | $k_3 \times 0.9$ | $\eta \times 1.1$ | $\eta \times 0.9$ |
|---|---|---|---|---|---|---|---|
| Year of Peak | 2081 | 2071 | 2062 | 2093 | 2088 | 2094 | 2089 |
| Difference from Model |  | 10 years | 19 years | 12 years | 7 years | 13 years | 8 year |